\documentclass[aps,prc,reprint,groupedaddress,showkeys,showpacs]{revtex4-2}
\usepackage{graphicx}
\usepackage{dcolumn}
\usepackage{bm}%
\usepackage{amsmath}
\usepackage[caption = false]{subfig}
\begin{document}


\title[]{Acceleration of charged particles in rotating magnetized star}

\author{Debojoti Kuzur}

\author{Rupamoy Bhattacharyya}%

\author{Ritam Mallick}
\email{mallick@iiserb.ac.in}
\homepage{\\http://home.iiserb.ac.in/~mallick/index.html}
\affiliation{Indian Institute of Science Education and Research Bhopal, Madhya Pradesh 462066, India}%

\date{\today}

\begin{abstract}
Charged particles at the crust of compact stars may be ejected and accelerated by the electric field
generated due to the rotation of the magnetized star. For neutron or hybrid
stars, the negatively charged particles are usually electrons, and the positively charged particles are mainly protons and Iron. Whereas the existence of
strange stars also includes the possibility of ejection of strangelets from the star surface. The flux of such strangelets emitted from all known pulsars is in the range of $10^9-10^{10}$ GeV. Therefore, such massive strangelets can be one of the candidates for the sources of the highest-energy cosmic rays that have still eluded us. Our model proposes a possible origin of these ultra high energy cosmic rays.
\end{abstract}

\maketitle

\section{Introduction}

The conjecture that strange quark matter (SQM) is the actual ground state of quantum chromodynamics (QCD) at high enough density/temperature is still in debate \cite{bodmer,witten}. Strange quark matter consists of an almost equal number of up, down, and strange quarks. The confirmation still awaits both from the theoretical and experimental regimes. In the experimental front collider, experiments are trying to confirm the existence of quark-gluon plasma at high temperatures. However, in the high-density regimes, no significant progress has been achieved in the earth-based experiments, and the only avenue that examines such a regime is from the astrophysical observation. If the conjecture is true, then the strange matter can exist in the form ranging from strangelets or strange nuggets to strange stars (SS). Strangelets are lumps of strange matter with baryon number (A) less than $10^6 - 10^7$. The limit is obtained from theoretical calculation \cite{witten,farhi}; however, the calculation is done with poorly determine parameters like strange quark mass, the bag constant, and strong coupling constant. The strangelets can be formed from various sources like high energetic nuclear collision \cite{greiner}, the collision of strange stars \cite{madsen2}, or during a supernova explosion \cite{vucetich}.

If such is the case, then a significant amount of strangelets should be present in cosmic rays. The search for strangelets in cosmic rays is a direct way of confirming the strange matter hypothesis. If the strange quarks stars do exist, then strangelets from the pulsar magnetosphere may populate the cosmic ray shower on earth. There has been previous calculation regarding the strangelets production in astrophysical scenarios. The important being one from Madsen \cite{madsen}, where he assumed that the strangelets generated from the collision of stars could reach the earth. He calculated the flux of strangelets in cosmic rays coming to earth and predicted the direct test for strange matter hypothesis. Cheng \& Usov \cite{cheng} argued that the strangelets could both be produced and accelerated from pulsars to reach us.   

The strangelets can also be one of the candidates for the high energy cosmic ray particles whose complete understanding still eludes us. The possible sources of energy of the particles near and beyond the ``ankle'' (i.e., energy $\sim 10^{9}$ GeV/particle) in the cosmic ray energy spectrum \cite{Fenu:2017hlc} remains unknown. Moreover, experimental uncertainty in their mass composition \cite{PhysRevD.99.022002} doesn't exclude the speculation on the existence of strangelets as well as \cite{Ren_Xin_2003}. 
In this regard, a novel approach of estimating the energy and flux (both near the origin and at the solar neighborhood) of some specimen particles (strangelets of $A=10^{2}$, $A=10^{4}$ and $A=10^{6}$ together with electron, proton, and Iron), that may escape from spinning, magnetized, relativistic stars - is described below. 

\section{Formalism}

In this paper, our source of strangelets is highly magnetized pulsars. The high magnetic field in rotating strange stars gives rise to a strong electric field at the surface. The electric field is strong enough to pull away charged particles from the surface of the star. In the following section, we describe our magnetized star and calculate the electric field at the star surface and find the energies of the particles (electron, proton, Iron, and strangelets) ejected from the star surface. 

\subsection{Magnetic field in a Magnetar}
We start with a spherically symmetric star coupled to a poloidal magnetic field $B$. The metric for such a star is described by
\begin{equation}
    ds^2=-e^{\nu(r)} dt^2+e^{\lambda(r)}dr^2+r^2 d\theta^2+r^2\sin^2\theta d\phi^2.
    \label{spherical}
\end{equation}
Maxwell equation for a given current density $J^\nu$ is
\begin{eqnarray}
\partial_\mu F^{\mu\nu}=J^\nu \\
\Rightarrow \partial_\mu (g^{\alpha\mu}g^{\beta\nu}F_{\alpha\beta})=g^{\lambda\nu}J_{\lambda}
\end{eqnarray}
The axisymmetric current density for the star is in the direction of $\phi$ and of the following form
\begin{equation}
    J_\mu=(0,0,0,J_\phi)
\end{equation}
which gives rise to an axisymmetric vector potential
\begin{equation}
    A_{\mu}=(0,0,0,A_\phi).
    \label{vecpot}
\end{equation}
Both the current $J_\phi$ and the vector potential $A_\phi$ in general are functions of both $r$ and $\theta$. These functions are separated by expanding them in terms of Legendre polynomials $P_n(\cos\theta)$
\begin{eqnarray}
    A_\phi=a_1(r)\sin{\theta}\frac{dP_1(\cos\theta)}{d\theta}, \\
    J_\phi=j_1(r)\sin{\theta}\frac{dP_1(\cos\theta)}{d\theta}.
\end{eqnarray}
The Maxwell equation for such a magnetar is calculated from the equation
\begin{equation}
    \frac{d^2a_\phi}{dr^2}+\frac{(\nu'-\lambda')}{2}\frac{da_\phi}{dr}=-4\pi e^\lambda j_\phi,
    \label{max}
\end{equation} 
where $a_\phi\equiv a_1$ and $j_\phi\equiv j_1$ are the coefficients for the expansion of $A_\phi$ and $J_\phi$. Also the magnetic field components are given by
\begin{equation}
    B^i=\epsilon^{ijk}\partial_j A_k.
\end{equation}
As we have taken an axisymmetric poloidal magnetic field, only $B_r$ and $B_\theta$ component will survive \cite{konno}, which are
\begin{eqnarray}
B_r=\frac{2\cos\theta}{r^2}a_\phi, \\
B_\theta=-\frac{e^{-\frac{\lambda}{2}}\sin\theta}{r}a'_\phi.
\end{eqnarray} 

The Maxwell equations (\ref{max}) is solved numerically by taking the initial conditions
\begin{equation}
    a_\phi\approx\alpha_0r^2
\end{equation} and 
\begin{equation}
    j_\phi=c_0r^2(\rho_0(r)+p_0(r))
\end{equation}
at the center of the star, where $\rho_0$ and $p_0$ are the density and pressure of the spherical unperturbed star. The values of $\alpha_0$ and $c_0$ determines the strength of magnetic field.

\begin{figure}
    \centering
    \includegraphics[scale=0.6]{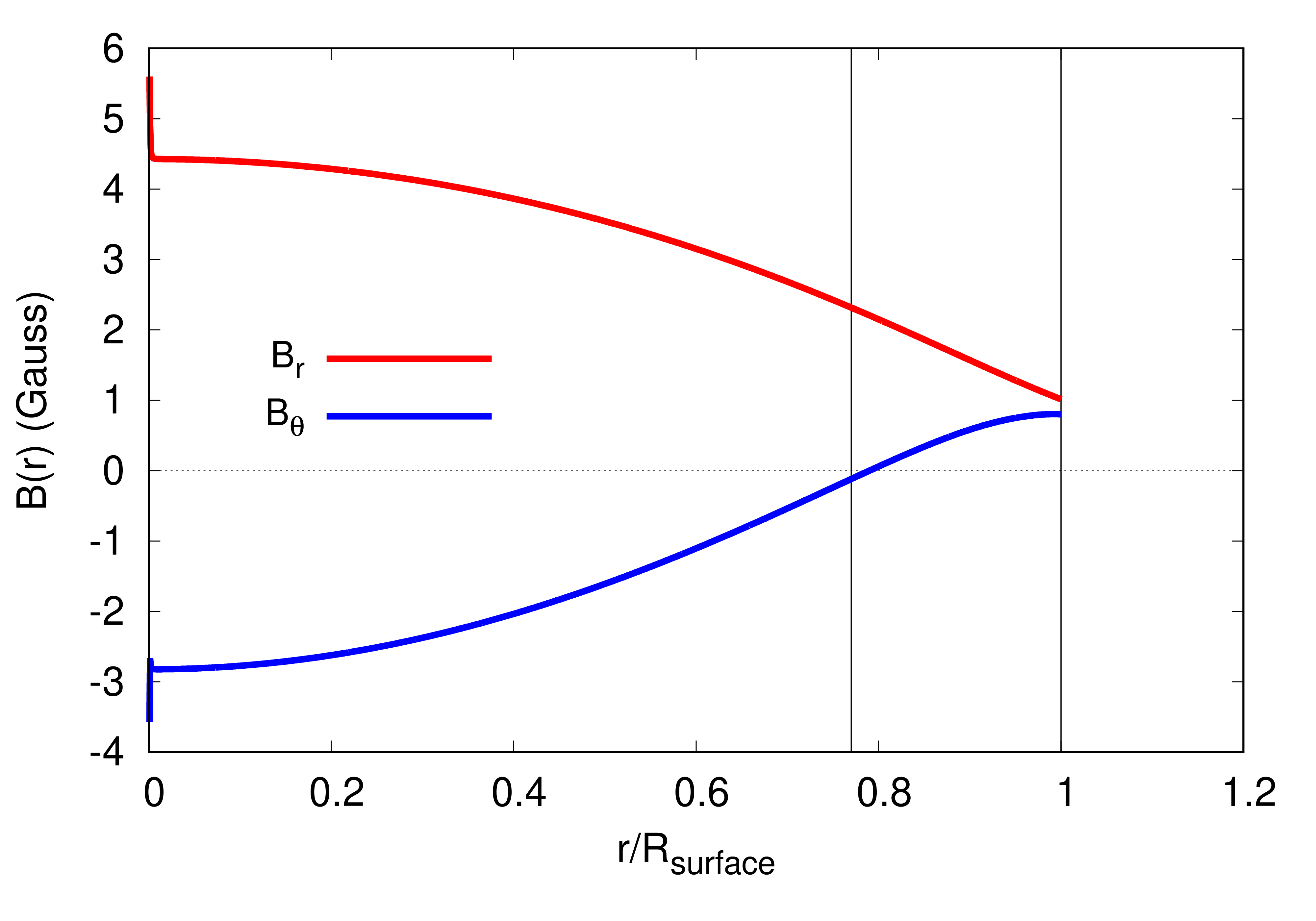}
    \caption{Plot showing the variation of $B_r$ and $B_\theta$ as a function of $r$. $B_r$ is marked with red line and $B_\theta$ with blue line.}
    \label{figBr}
\end{figure}

In figure \ref{figBr}, $B_r$ and $B_\theta$ has been plotted as a function of $r$. It can be seen that though the $B_r$ component remains positive entirely from the center to the surface of the star, the $B_\theta$ component of magnetic field changes sign at approximately $0.75R_{s}$. This will impact on the nature of the charge (positive or negative) of particles that can be ejected from the magnetar, as will be discussed in the further sections.

\subsection{Structure of the Magnetar}
We start with a spherically symmetric star and add a poloidal, axisymmetric magnetic field to it. This magnetic field distorts the star and makes it an oblate spheroid \cite{konno}, and the metric is given by (\ref{spherical}),
\begin{eqnarray}
ds^2\Big|^M=-e^{\nu(r)}[1+2(h_0^M(r)+h_2^M(r)P_2(\cos\theta))]dt^2 \\ \nonumber
 +e^{\lambda(r)}\Big[1+\frac{2e^{\lambda(r)}}{r}(m_0^M(r)+m_2^M(r)P_2(\cos\theta))\Big]dr^2 \\ \nonumber
 +r^2[1+2k_2^M(r)P_2(\cos\theta)](d\theta^2+\sin^2\theta d\phi^2)
\end{eqnarray}


where $h_0^M,h_2^M,m_0^M,m_2^M,k_2^M$ are the coefficients of Legendre polynomials and the superscript $' M'$ stands for magnetic field perturbation.

For a rotating magnetar, we include the frame dragging effect around the rotating star \cite{hartle}. In our calculation, we consider an aligned rotator, that is, the magnetic axis and the rotation axis is aligned to each other. The frame dragging velocity $\omega(r,\theta)$ is the velocity with which the inertial coordinates around the star is dragged. 
 The rotation will again incorporate further centrifugal distortions to the star. The metric for the rotating star is given by
 \begin{eqnarray}
 ds^2\Big|^R=-e^{\nu(r)}[1+2(h_0^R(r)+h_2^R(r)P_2(\cos\theta))]dt^2 \\ \nonumber
 +e^{\lambda(r)}\Big[1+\frac{2e^{\lambda(r)}}{r}(m_0^R(r)+m_2^R(r)P_2(\cos\theta))\Big]dr^2 \\ \nonumber
 +r^2[1+2k_2^R(r)P_2(\cos\theta)](d\theta^2+\sin^2\theta (d\phi-\omega(r,\theta)dt)^2).
\end{eqnarray}

The superscript $'R'$ stands for rotational perturbation. Hence as the system is an aligned magnetic rotator, all the properties of the star, such as mass and ellipticity can be obtained linearly, that is adding the individual mass change and radius change of the star due to magnetic and rotational contributions. First we solve Einstein equation to find the equation of motion for the metric and then numerically solving for all the unknown functions $'M'$ and $'R'$. The mass of the star is then related as
\begin{equation}
    M_{t}=M_{sph}+\Delta M_{M}+\Delta M_{R},
\end{equation}
where $M_t$ is the total mass and
\begin{eqnarray}
\Delta M_{M}=m_0^M(R_s)+f(\mu,M_{sph},R_{s}) \\
\Delta M_{R}=m_0^R(R_{s})+f(\Omega,R_{s})
\end{eqnarray}
are the magnetic and rotational mass shifts, and $f$ is some function of $M_{sph}$ and $R_{s}$ which are the mass and radius of the spherical star respectively, and $\mu$ and $\Omega$, the magnetic moment and the total angular velocity of the star respectively.

Similarly, the radius can be written as
\begin{equation}
    R_{t}(\theta)=R_{s}+\Delta R_{M}(\theta)+\Delta R_{R}(\theta)
\end{equation}
where $R_t$ is the total radius and $\Delta R_{M}$ and $\Delta R_{R}$ are the radius corrections due to magnetic distortion and rotational distortions respectively. The explicit form is given in the Appendix.
The ellipticity is defined as
\begin{equation}
\epsilon \equiv\Big[1-\frac{R_{t}(0)}{R_{t}(\frac{\pi}{2})}\Big]^{\frac{1}{2}}
\end{equation}

\subsection{Induced electric field in a magnetar}
 As the magnetar is rotating with a poloidal magnetic field, the four vector potential (\ref{vecpot}) becomes
 \begin{equation}
     A_{\mu}=(A_t,0,0,A_\phi),
     \label{rotvecpot}
 \end{equation}     
where $A_t$ and $A_\phi$ are related to each other by
\begin{equation}
    A_t=-\Omega A_\phi.
    \label{ataphi}
\end{equation}
$A_t$ can be described as a scalar potential such that it gives rise to an electric field
\begin{equation}
    E_i=\partial_iA_t.
\end{equation}
Using eqn. (\ref{rotvecpot}) in the Maxwell equation eqn. (\ref{max}) and combining with eqn. (\ref{ataphi}) gives electric field inside the star in terms of known quantities \cite{konno2},
\begin{equation}
    E_{in}(r,\theta)=e^{-\frac{\lambda+\nu}{2}}\bar{\omega}a_\phi'\sin^2\theta.
    \label{electric_in}
\end{equation}

However, the more useful quantity in our case would be the electric field outside the surface of the star which is obtained by solving (\ref{max}) for (\ref{rotvecpot}) analytically and is given by

    $$E_{out}(r,\theta)=\frac{1}{2M^6r^3}\Big[c(J,\mu,M,r)\Big(4M^5r(6r^2-3Mr-M^2)$$
    $$+6M^4r^3(2r-3M)ln(1-\frac{2M}{r})\Big)-2J\mu M(24r^3-12Mr^2$$
    \begin{equation}
        -4M^2r-3M^3)-3J\mu r(8r^3-12Mr^2-M^3)ln(1-\frac{2M}{r})\Big]P_2(\cos\theta),
        \label{electric_out}
    \end{equation}
where $c(J,\mu,M,r)$ is a known function of the angular momentum, dipole moment, total mass and radius of the star.
 
\subsection{Ejection of particles from the surface}
\begin{figure}
    \centering
    \includegraphics[scale=0.6]{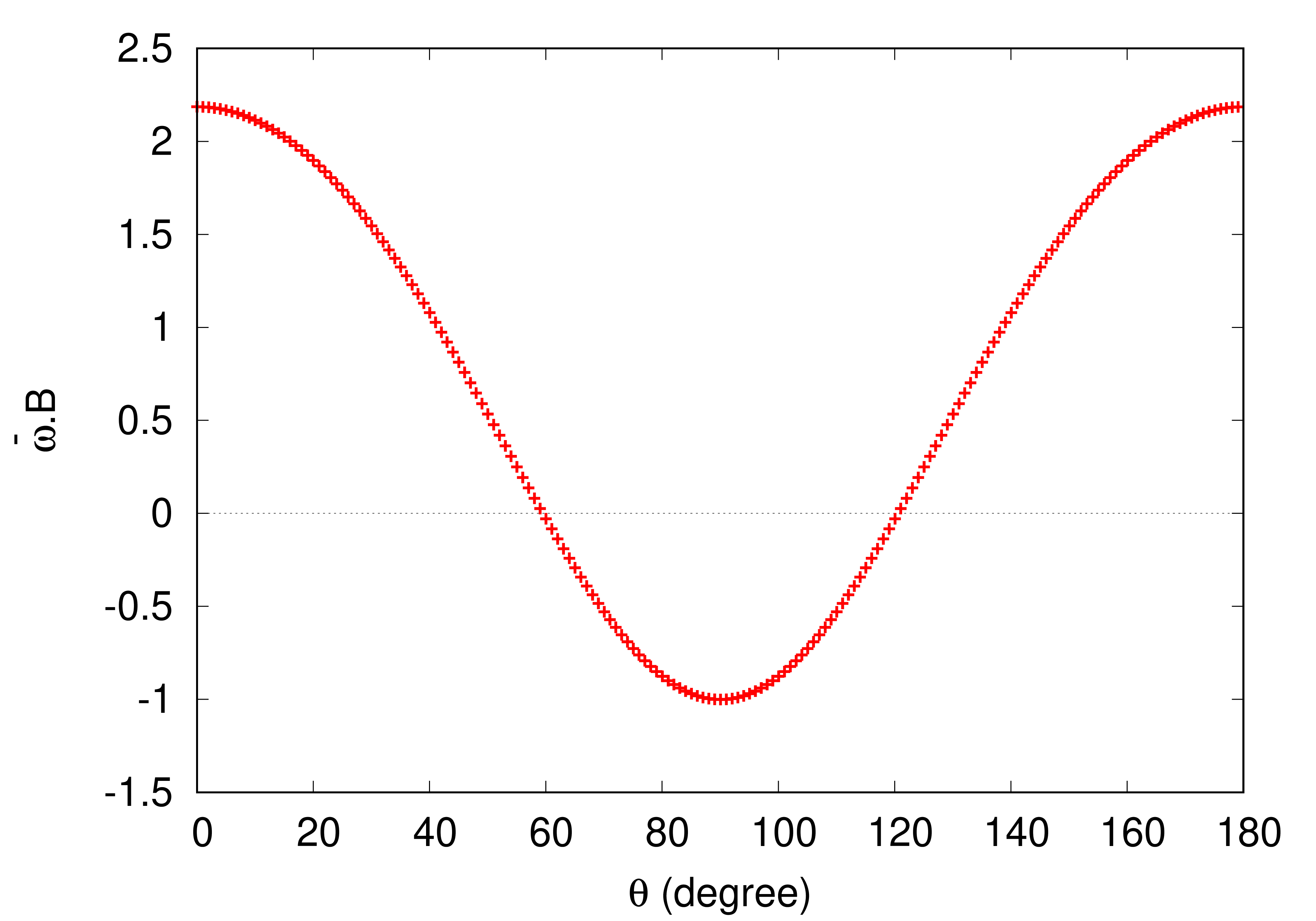}
    \caption{$\bar{\omega}\cdot B$ is plotted with respect to $\theta$, which is seen to remain positive except in the region $\frac{\pi}{3}<\theta<\frac{2\pi}{3}$}
    \label{omegaB}
\end{figure}

The electric field outside the star's surface drives particles on the surface out of the star. Charged particles on the star surface are dragged along with velocities
 \begin{equation}
     v^i=\epsilon^{ijk}\bar{\omega}_jx_k,
 \end{equation}
 where $\bar{\omega}_j$ is the general frame dragging velocity vector. In an electromagnetic equilibrium condition, we can write
 \begin{equation}
     E_i=-\epsilon_{ijk}v^jB^k.
     \label{lorentz}
 \end{equation}
Now from Maxwell equation we have
\begin{equation}
    \partial^i E_i=\frac{q}{\varepsilon_0},
\end{equation}
where $q$ is the charge of the particle that is ejected from the surface of the star. Solving for $q$, we have 

\begin{equation}
    q=2\varepsilon_0\bar{\omega_k}B^k=2\varepsilon_0\bar{\omega}\cdot B.
    \label{omegab}
\end{equation}
The quantity $\bar{\omega}\cdot B$ has been plotted in the figure \ref{omegaB} as a function of $\theta$ on the surface of the star. It could be seen that around the equator ($\frac{\pi}{3}<\theta<\frac{2\pi}{3}$), $\bar{\omega}\cdot B<0$. Hence from equation (\ref{omegab}) positively charged particles will be ejected from the surface around the equator. Also around the pole $\bar{\omega}\cdot B > 0$ and therefore negatively charged particles will be emitted from there.

\subsection{Energies of the particles}
Energies of the particle on the surface can be calculated by integrating the electric field (\ref{electric_out}) from infinity to the radius of the star. This energy will itself be a function of $\theta$
\begin{equation}
E(\theta)=\int_{\infty}^{R_{s}(\theta)}E_{out}.dr
\end{equation}

We have considered primarily two types of particles, hadronic and quark, that can be ejected from the surface of the star. The kinetic energy ($T$) with which the particles are ejected is
\begin{equation}
    T_{(h/q)}(\theta)=E(\theta)-m_{(h/q)}c^2-BE_{(h/q)}
\end{equation}
where subscript $h/q$ stands for "hadron or quark" and $m_{(h/q)}$ is the mass of the particles being ejected and $BE_{(h/q)}$ is the total binding energy of this particles on the star's surface. The electric field inside the star (eqn (\ref{electric_in})) is numerically solved for a given magnetic field and angular velocity of the star and thus the magnetic moment $\mu$ and angular momentum $J$ of the star is fixed. These values are passed outside the star for solving the electric field outside (eqn. (\ref{electric_out})). 
\begin{table*}
\caption{\label{table1} Highest kinetic energies of the nuclear particles ejecting out of the compact stars for different $B_{surf}$ and $\Omega$ combination is tabulated below using NL3 EoS }
\begin{ruledtabular}
\begin{tabular}{cccccccc}
 $B_{surf}$&$\Omega$&$M$&$\Delta_{Mass}$&$Ellipticity$&$T_e$&$T_p$&$T_{Fe}$ \\
 (G)&(Hz)&($M_{\odot}$)&($M_{\odot}$)&&(GeV)&(GeV)&(GeV) \\
 \hline
 &1003&1.66&0.25&0.31&5.06E+04&2.36E+04&6.15E+05 \\
 $\approx10^{14}$&525&1.41&0.01&0.17&2.76E+04 & 1.35E+04& 3.51E+05 \\
 &48&1.40&0.00&0.02&5.06E+04 & 2.36E+04 & 6.15E+05 \\
\hline
&1003&1.66&0.26&0.32&5.04E+05 & 2.34E+05& 6.09E+06 \\
$\approx10^{15}$&525&1.42&0.02&0.2&2.75E+05 & 1.34E+05& 3.48E+06\\
&48&1.41&0.00&0.1&2.46E+04& 1.22E+04& 3.18E+05 \\
\hline
&1003&2.01&0.61&0.72&3.94E+06 & 1.25E+06& 3.24E+07 \\
 $\approx10^{16}$&525&1.77&0.37&0.71&2.13E+06& 6.84E+05& 1.78E+07 \\
 &48&1.76&0.36&0.1&1.91E+05& 6.24E+04& 1.62E+06 \\
\end{tabular}
\end{ruledtabular}
\end{table*}

\begin{table*}
\caption{\label{table2} Highest kinetic energies of the ordinary strangelets of different masses ejecting out of the compact stars for different $B_{surf}$ and $\Omega$ combination is tabulated below. Here charges of the ordinary strangelets having different masses are calculated using Ref. \cite{PhysRevD.71.014026} using strange matter EoS}
\begin{ruledtabular}
\begin{tabular}{cccccccc}
 $B_{surf}$&$\Omega$&$M$&$\Delta_{Mass}$&$Ellipticity$&$T_{A=10^2}$&$T_{A=10^4}$&$T_{A=10^6}$ \\
 (G)&(Hz)&($M_{\odot}$)&($M_{\odot}$)&&(GeV)&(GeV)&(GeV) \\
 \hline
 &1003&1.28&0.02&0.26&5.78E+04 &1.28E+06  &2.85E+07 \\
 $\approx10^{14}$&525&1.27&0.00&0.14&3.16E+04 &7.03E+05   &1.60E+07 \\
 &48&1.26&0.00&0.02&2.99E+03 &7.43E+04 &2.47E+06 \\
\hline
&1003&1.28&0.02&0.27&5.74E+05   &1.26E+07  &2.74E+08 \\
$\approx10^{15}$&525&1.27&0.00&0.16&3.13E+05 &6.90E+06   &1.50E+08\\
&48&1.27&0.00&0.08&2.87E+04   &6.41E+05 &1.47E+07 \\
\hline
&1003&1.47&0.21&0.68&3.97E+06   &8.73E+07  &1.89E+09\\
 $\approx10^{16}$&525&1.46&0.19&0.66&2.14E+06   &4.71E+07   &1.02E+09 \\
 &48&1.46&0.19&0.66&1.95E+05   &4.30E+06  &9.38E+07\\
\end{tabular}
\end{ruledtabular}
\end{table*}

\begin{table*}
    \caption{\label{table3} Kinetic energies of particle ejection from real pulsars with observed magnetic field and rotational frequencies using NL3 EoS}
\begin{ruledtabular}
    
    \begin{tabular}{cccccc}
    $Name$&$B_{surf}$&$\Omega$&$T_p$&$T_{Fe}$&$T_{Str(A=10^6)}$ \\
    &(G)&(Hz)&(GeV)&(GeV)&(GeV) \\
    \hline
    $PSRJ1748-2446ad$&$\approx10^{14}$&716.56&1.05E+00&5.74E+01&1.00E+06 \\
    
    $SGR 1806-20$&$\approx2\times10^{15}$&7.55&1.11E+03&2.88E+04&3.10E+06 \\
    \end{tabular}

\end{ruledtabular}
\end{table*}

\begin{figure}[!htb]
    \centering
    \centering
    \includegraphics[scale=0.7]{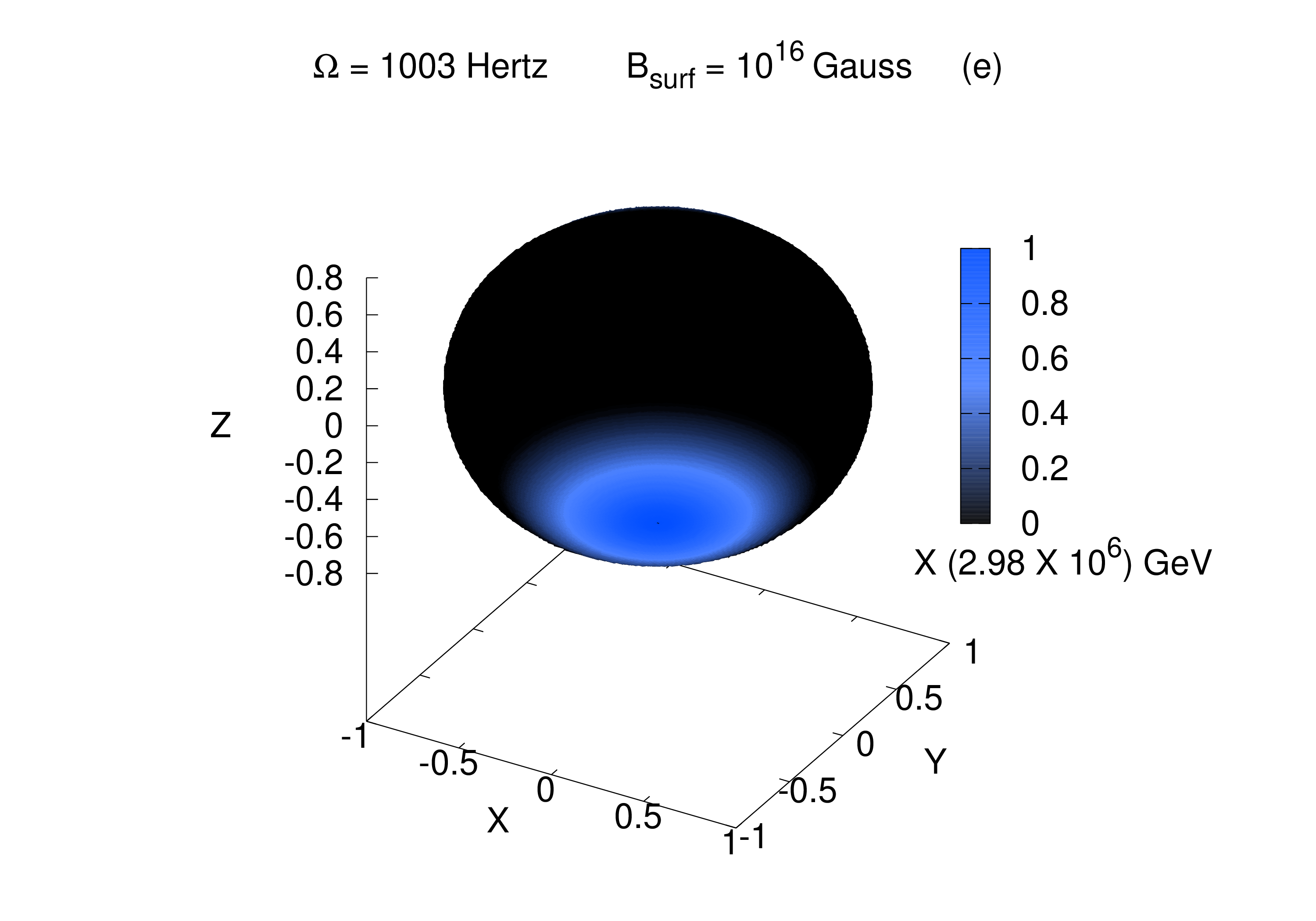}
    \centering

    \includegraphics[scale=0.7]{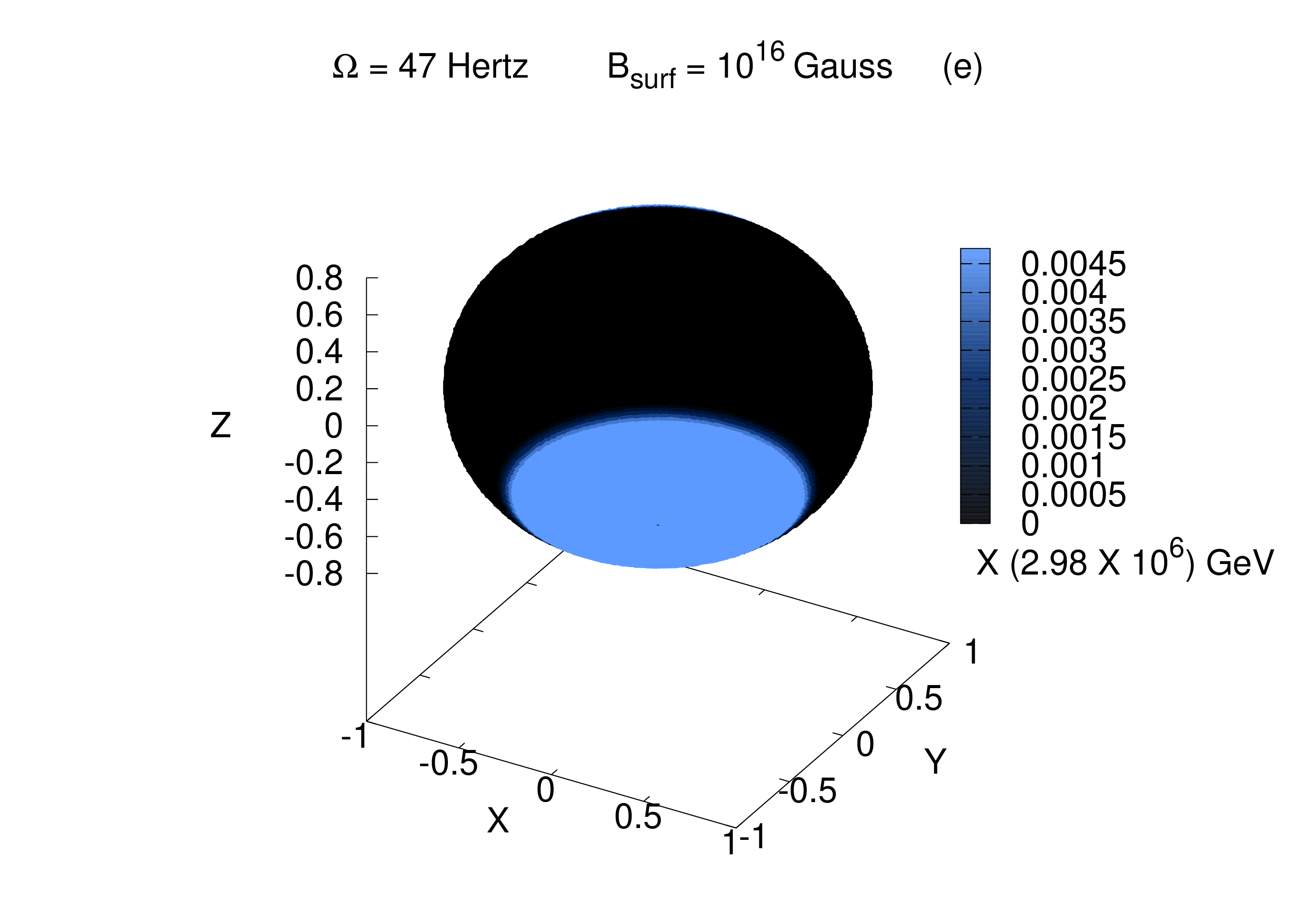}
    \caption{Kinetic Energy of electrons ejecting from the polar regions of neutron stars with magnetic field $B_{surf}\approx10^{16}$ G and different angular velocities.}
    \label{plot1}
\end{figure}

\begin{figure}[!htb]
    \centering
    \centering
    \includegraphics[scale=0.7]{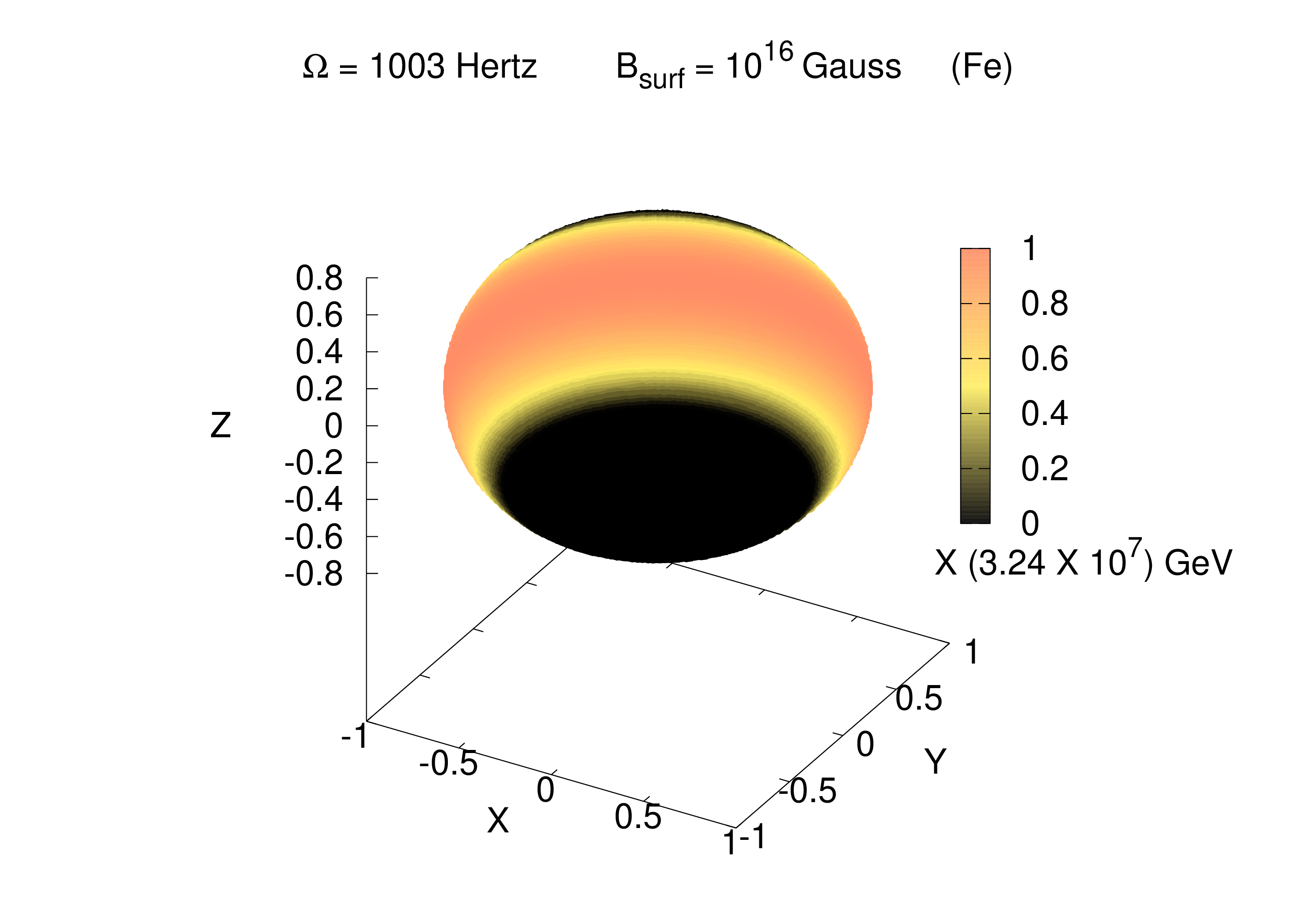}
    
    \centering

    \includegraphics[scale=0.7]{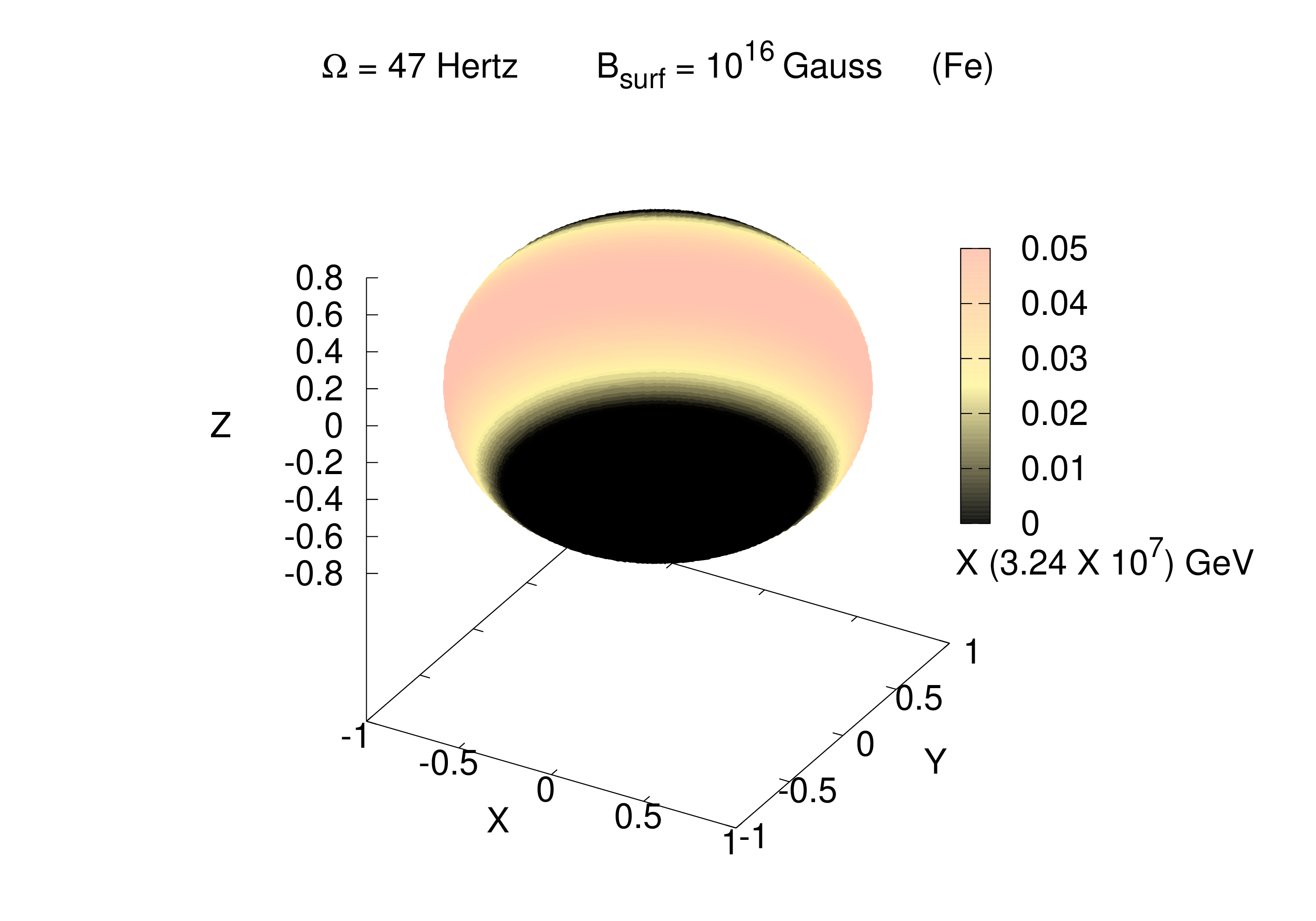}
    \caption{Energy plots of the equatorial ejection of iron  from neutron star with magnetic field $B_{surf}\approx10^{16}$ G and different angular velocities.}
    \label{plot2}
\end{figure}

\begin{figure}[!htb]
    \centering

        \centering
        \includegraphics[scale=0.7]{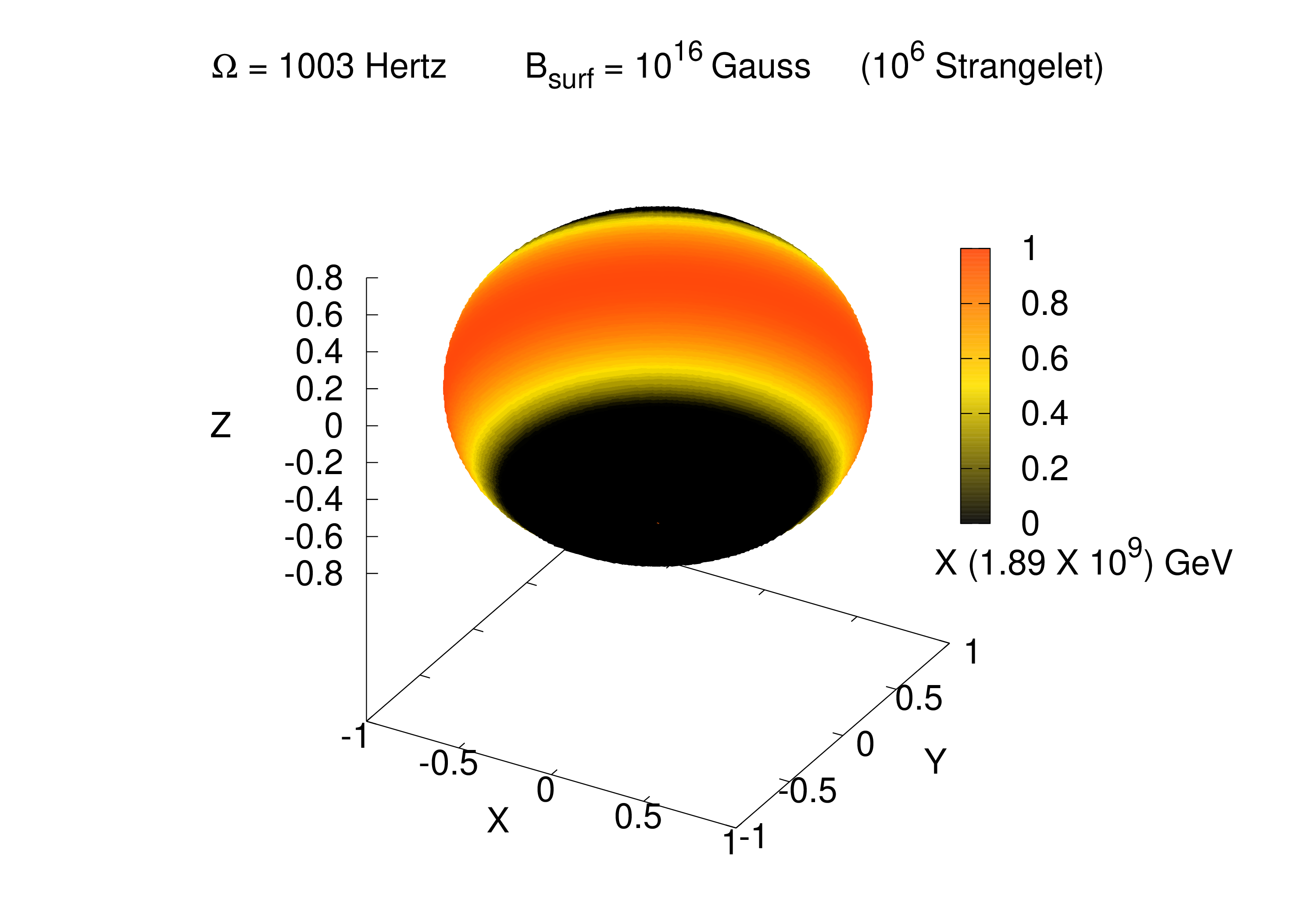}


        \centering

    \includegraphics[scale=0.7]{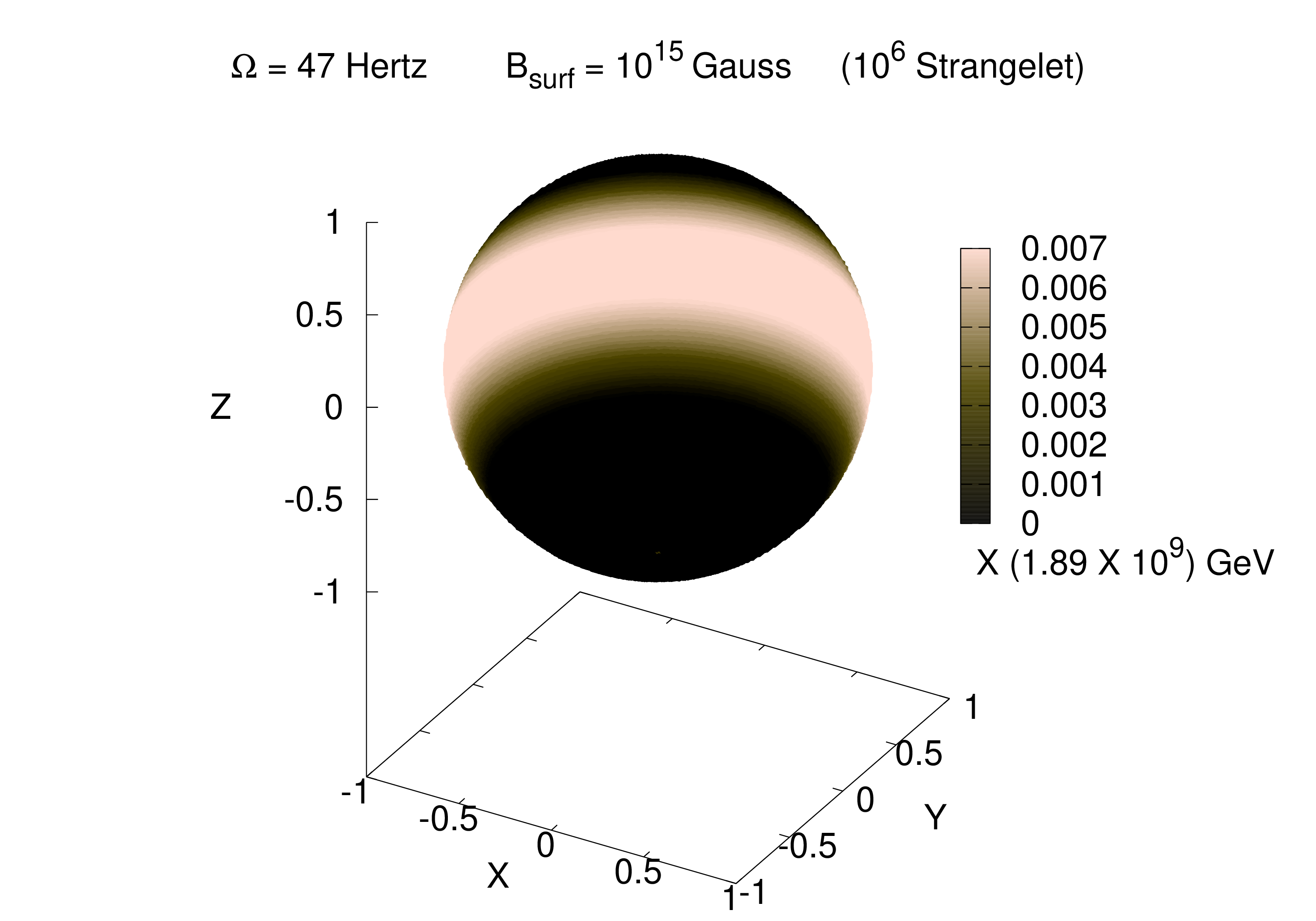}
    \caption{Energy plots of the equatorial ejection of strangelets from neutron star with magnetic field $B_{surf}\approx10^{16}$ G and different angular velocities.}
    \label{plot3}
\end{figure}

\section{Results}

We first calculate the total $T$ for electron, proton, Fe (Iron),   and strangelet (with A$= 10^2$, $10^4$, $10^6$). 
For calculating the energy and flux of electron, proton, and neutron, we have used NL3 EoS \cite{nl3} to construct a neutron star whereas for calculating the energy and flux of strangelet we have used bag model EoS \cite{chodos,alford,weiss} to construct the strange star. The SM has up and down quarks with masses $5$ MeV and strange quarks with mass $100$ MeV. The bag constant is assumed to be $B^{1/4}=140$ MeV, and the strength of quark coupling is $a_4=0.5$.
Table \ref{table1},\ref{table2},\ref{table3} shows the highest kinetic energy of different particles emitted from the star surface, and the figures give the heat map of the particles emitted from a different region of the star.

In Table \ref{table1}, we have given the Kinetic energy of the particles that are emitted out of the star surface. The particles shown in the plots are the electron, proton, and Iron, emitted from the surface of stars having three different surfaces magnetic field. The stars having the same surface magnetic field can have different rotational speeds. Therefore, we have taken three different rotational speed of stars. We find that as the magnetic field increases (which thereby increases the surface electric field), the kinetic energy emitted from the star surface increases. The kinetic energy also increases with an increase in rotational speed. 
Also seen from Table \ref{table1} is that as the particle mass increases (for a particular charge like proton and Iron), the kinetic energy of emitting particles increases. 
The electron is a negatively charged particle that is emitted from the polar region, whereas the positively charged particles (like proton, Iron, and strangelets) are emitted from the equatorial region. Although electrons have small mass, the kinetic energy of the electron is higher than proton because the electric field at the poles is higher than at the equator from where the electrons are emitted.

The observed isotropy of the cosmic rays of kinetic energies beyond $10^6$ GeV in the ground-based experiments confirms that the sources in this region are extra-galactic. Moreover, the composition of the cosmic rays at and above this energy is unknown as extensive air-shower technique can't measure the charge of the mass of the primary cosmic rays that impinge at the top of the atmosphere. So while estimating the kinetic energies in Table \ref{table1} for electron, proton, and Iron, we have incorporated the contributions of strangelets of different baryon numbers in Table \ref{table2}. 

In Table II, we show the kinetic energy of strangelets as they are emitted from the star surface. We have considered three different strangelets baryon number ($A=10^2$, $10^4$ and $10^6$). Similar to the previous Table, three different magnetic fields and three different rotational velocities are considered. The results show similar nature as that deduced from the previous Table. Also, as $A$, the kinetic energy of the emitted particles also increases.

The plots (basically heat maps) corresponding to the table (for the electron, iron and strangelet $A=10^6$) are given in fig \ref{plot1}, fig \ref{plot2} and fig \ref{plot3}. We find that most of the positive particles emitted from the star surface are more or less limited between $\frac{\pi}{3}<\theta<\frac{2\pi}{3}$. 
The highest energies for the strangelet emission could be seen for the first plot in fig \ref{plot2}, where the magnetic field and rotation are taken to be maximum.
All the other plots are normalized with the maximum energy.

Table III and Table IV show the kinetic energy of positive particles for two well-known pulsars (PSR J1748-2446ad and SGR 1860-20). We see that the strangelets emitted from such pulsars have kinetic energy a few times of $10^6$ GeV.

\subsection{Flux estimation}
All the energy that is emitted from the surface of the star does not reach us. To compare with the energetic particles which we detect in cosmic rays, we should find the flux of such particles that are reaching us.
The flux of the particle emission is calculated by defining the rate of ejection
\begin{equation}
\frac{dN}{dt}=A \times nc,
\end{equation} 
where $A$ is the area from which the particles are emitted and $n=\frac{q}{e}$ is the charge density, where $q$ is the total charge, and $e$ is the elementary charge of the particle and $c$ is the speed of light. The area element is
\begin{equation}
    dA=R_{t}(\theta)^2\sin\theta d\theta d\phi
\end{equation}
where the radius itself is a function of $\theta$. Hence total area is given by
\begin{equation}
A=2\pi\int_{\frac{\pi}{3}}^{\frac{2\pi}{3}}R_{t}(\theta)^2\sin\theta d\theta.
\end{equation}
The analytical expression for the total area is given in Appendix. Thus the particle ejection rate is now defined as
\begin{equation}
    \frac{dN}{dt}=A\frac{2\varepsilon_0\bar{\omega}.B}{e}c.
    \label{ndot}
\end{equation}

\begin{figure}[!h]
    \includegraphics[scale=0.6]{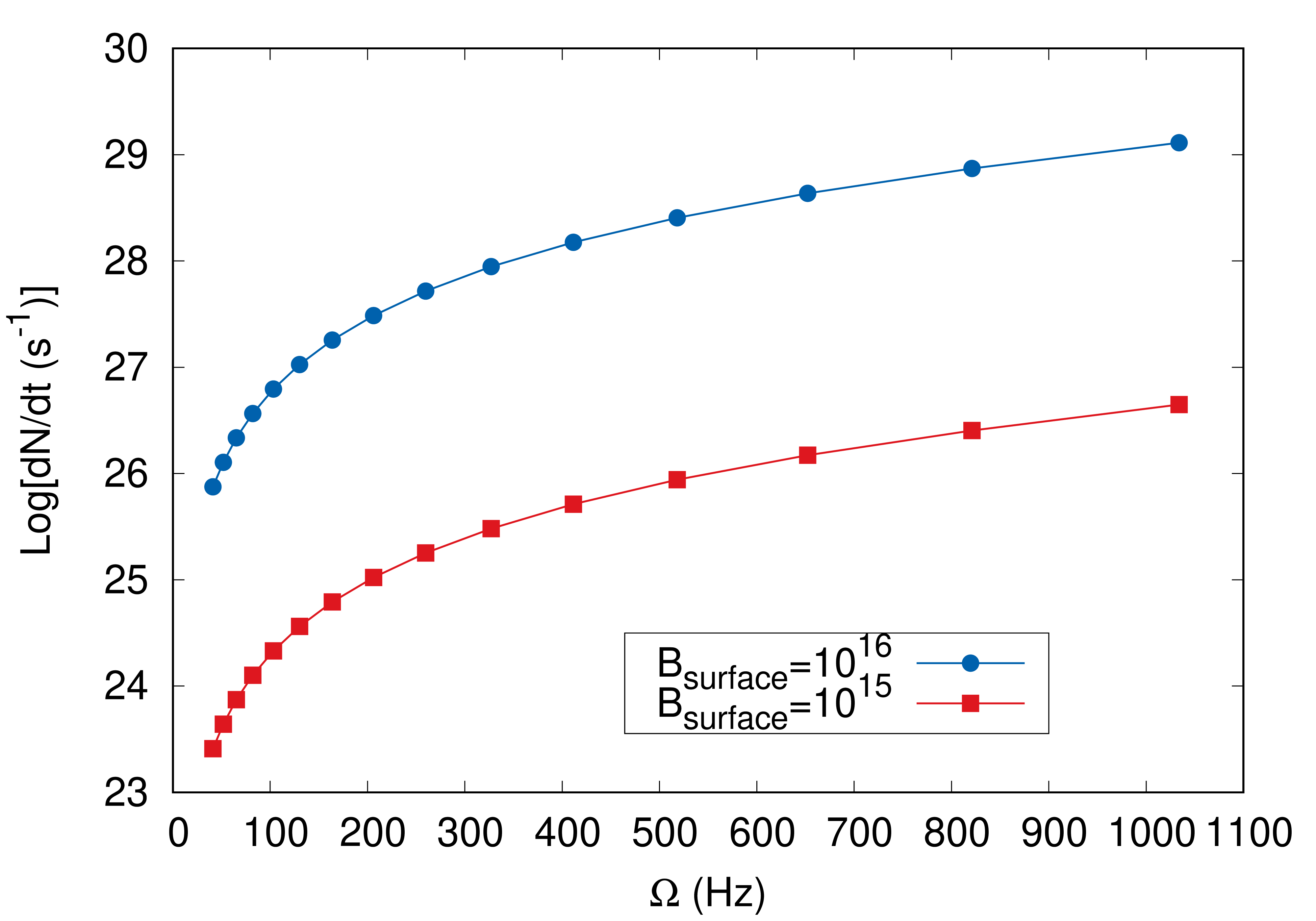}
    \caption{The rate of $A=10^6$ strangelet ejection, at different $B$ has been plotted as a function of $\Omega$.}
    \label{figndot}
\end{figure}

As the strangelets with $A=10^6$ has the maximum kinetic energy when it leaves the star surface, we would concentrate our focus on such massive strangelets.
The logarithm of the particle ejection rate as a function of rotational velocities is plotted in fig \ref{figndot}. We have plotted the ejection rate for two different surface magnetic field values. We find that the rate increases as the magnetic field strength increase and is also high for a rapidly rotating star. Thus strangelets emitted from different stars reach the earth's surface with different energy values and different flux.

To calculate the flux of these particles, we describe how these pulsars/magnetars are distributed around us. We take care of two aspects:
\begin{enumerate}
    \item distribution of pulsars with respect to distance from earth \cite{andreasyan}.
    \item distribution of pulsars with respect to rotational velocities and Magnetic field \cite{Pan}.\\
\end{enumerate}  

\begin{figure}[!h]
    \includegraphics[scale=0.6]{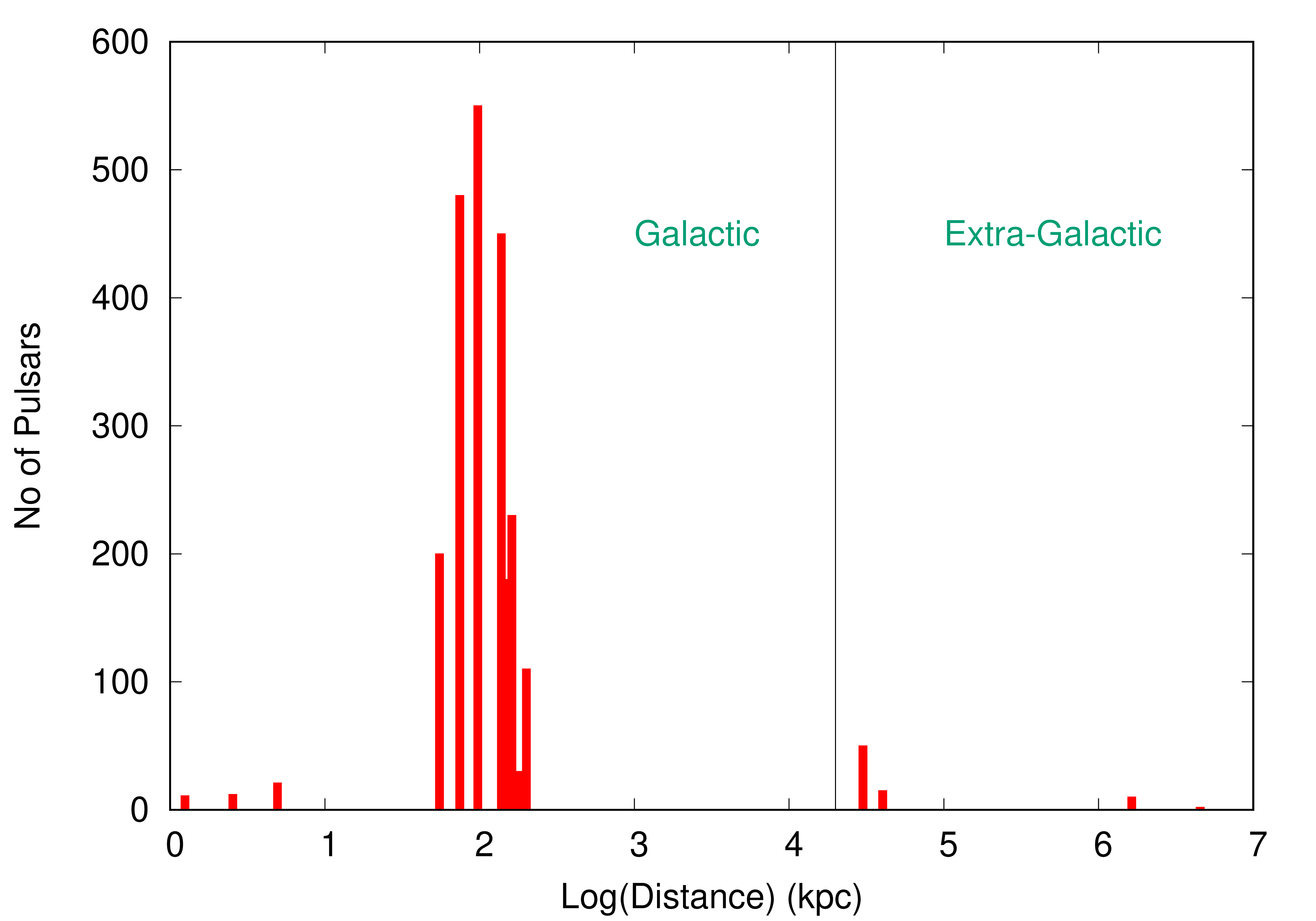}
    \caption{Distribution of pulsars with respect to distances from earth in kpc has been plotted close to the observed distribution, where the distances has been divided into galactic and extra-galactic regions.}
    \label{distance}
\end{figure}

As it is observed in cosmic rays, we assumed that the distance is divided into the galactic and extra-galactic region. Distances $\le$ 15 kpc has been taken to be galactic and extra-galactic distances are taken to be in the region 100 kpcs $\le$ and $\le$ 700 kpc.
A total of 2400 pulsars have been divided into these regions, and their distribution is shown in figure \ref{distance}. The distribution shows that most of the pulsars that we are able to record lies mostly in the galactic region, and a very few in the extra-galactic region.

\begin{figure}[!htb]
    \centering
    
    \centering
    \includegraphics[scale=0.6]{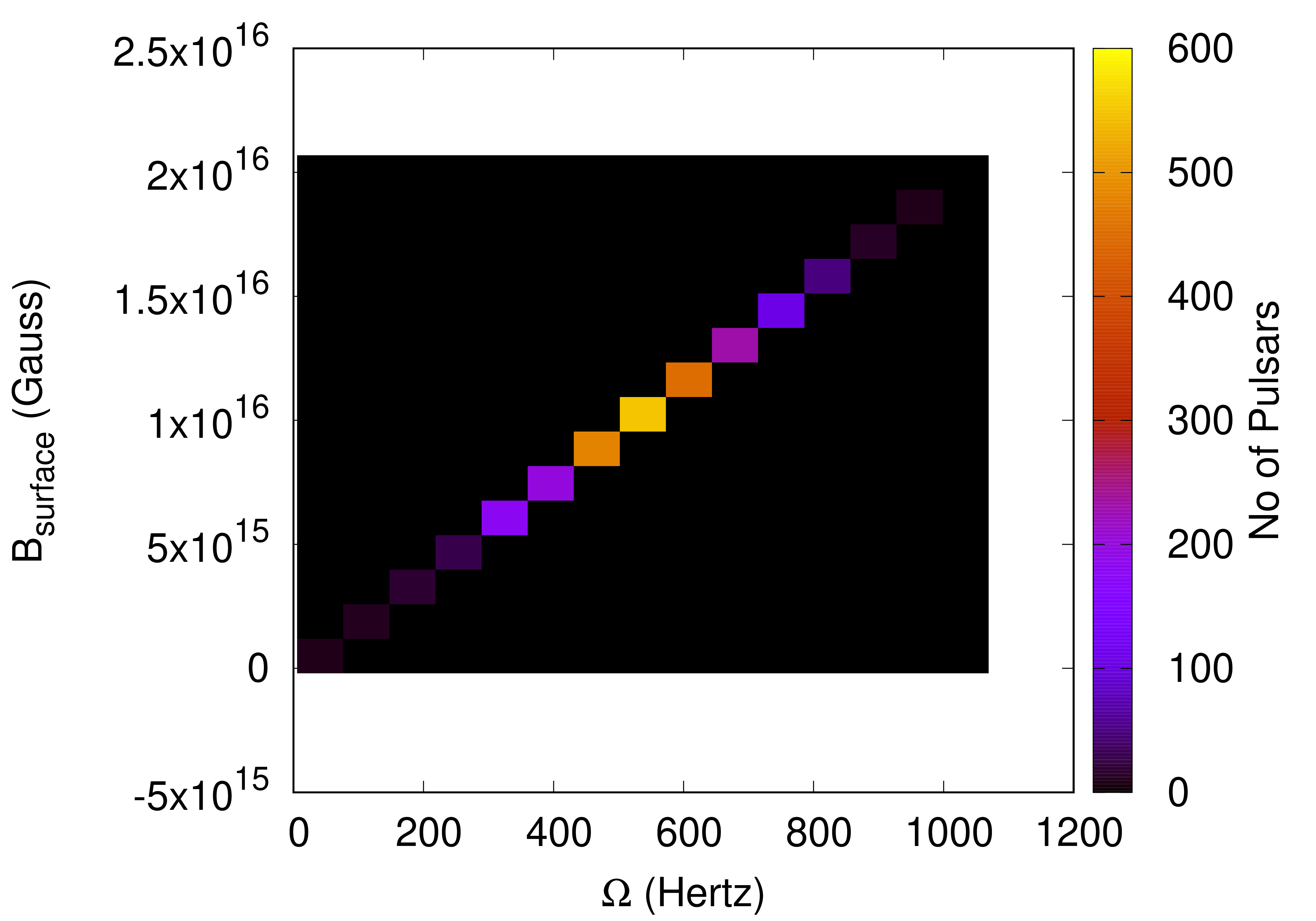}
    
    
    \centering
    
    \includegraphics[scale=0.6]{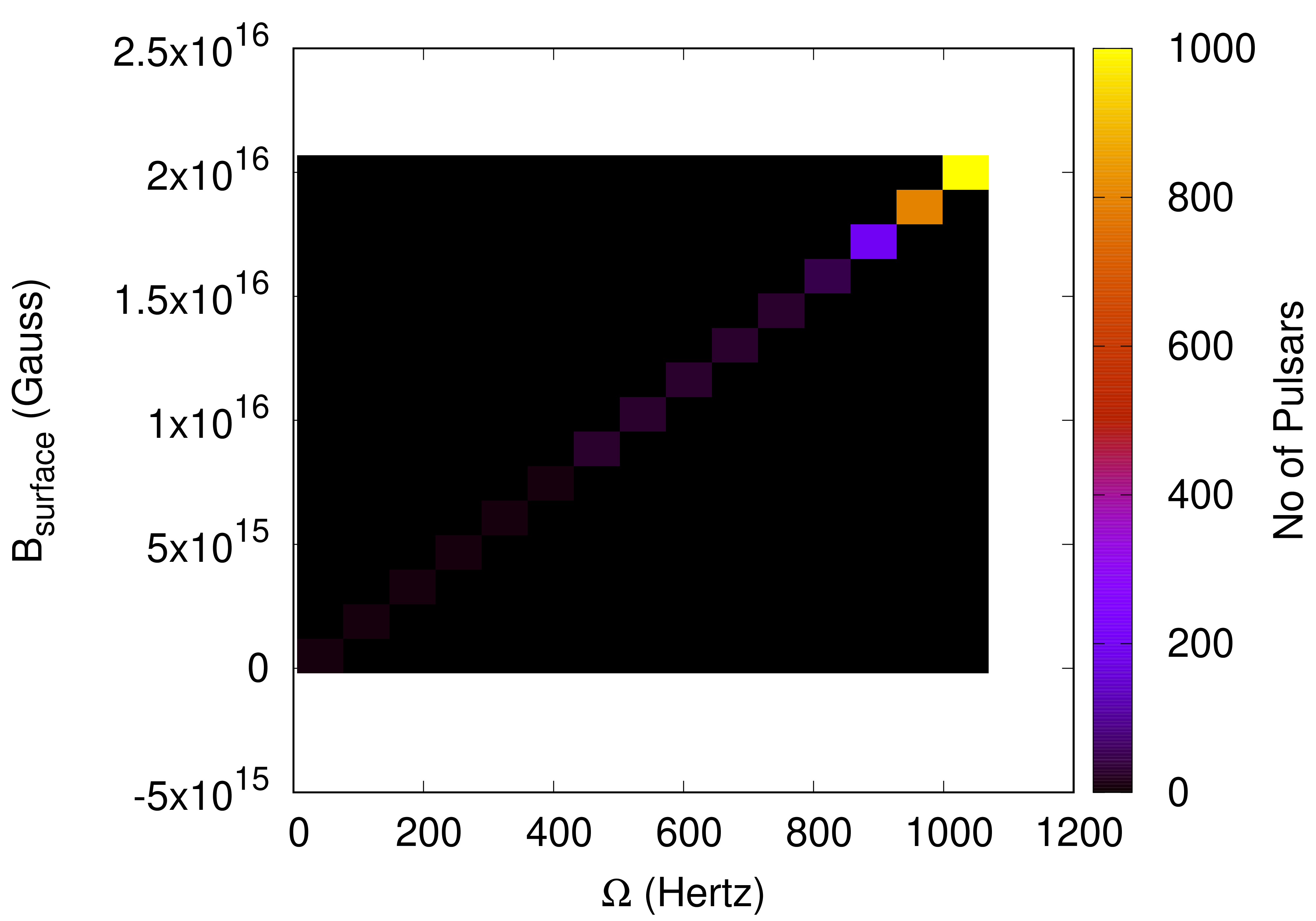}
    \caption{Density of pulsars with respect to $\Omega$ and $B_{surface}$ has been plotted approximated to the observed distribution in the first figure and an extreme case in the second figure where most of the pulsars are highly spinning and highly magnetized.}
    \label{dis}
\end{figure}

Each of these pulsars also has some rotational velocity and magnetic field. We have taken two such distributions, which are shown in figure \ref{dis}. The first of these distributions is the usual pulsar distribution \cite{Pan}, where most of them have a moderate magnetic field and rotational velocities. For comparison, we have also taken a second distribution which is an extreme case where almost all the pulsars have high rotation and high magnetic field values.\\

\begin{figure}[!htb]
    \centering
    \includegraphics[scale=0.6]{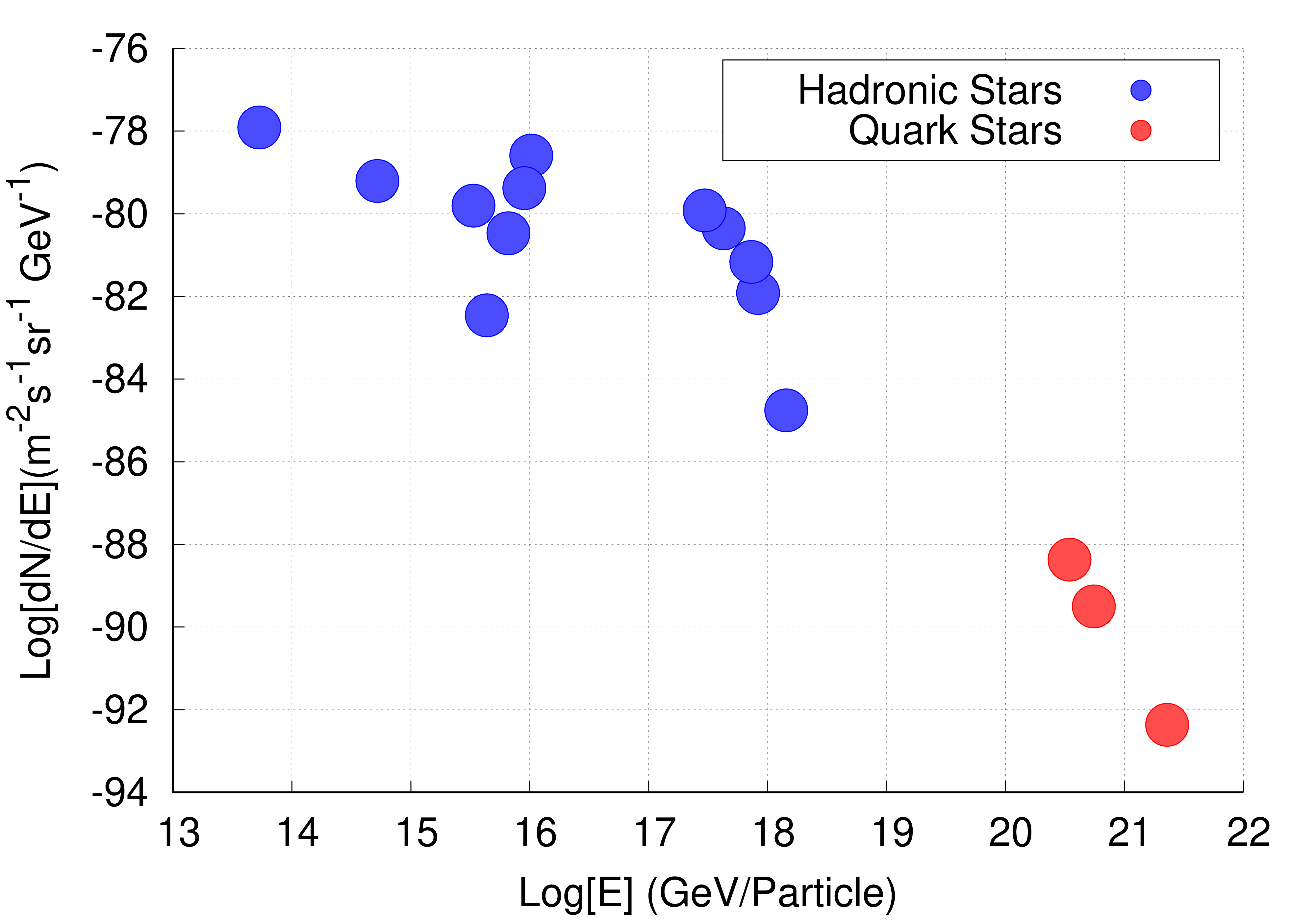}
    \caption{Log plot of the differential flux with respect to the energy per particle reaching earth. This flux obtained depends on both the distance distribution and the observed spin vs. magnetic strength distribution. Galactic pulsars are taken to be hadronic stars, but extra-galactic are taken to be quark stars.}
    \label{flux1}
\end{figure}

\begin{figure}[!htb]
    \centering
    \includegraphics[scale=0.6]{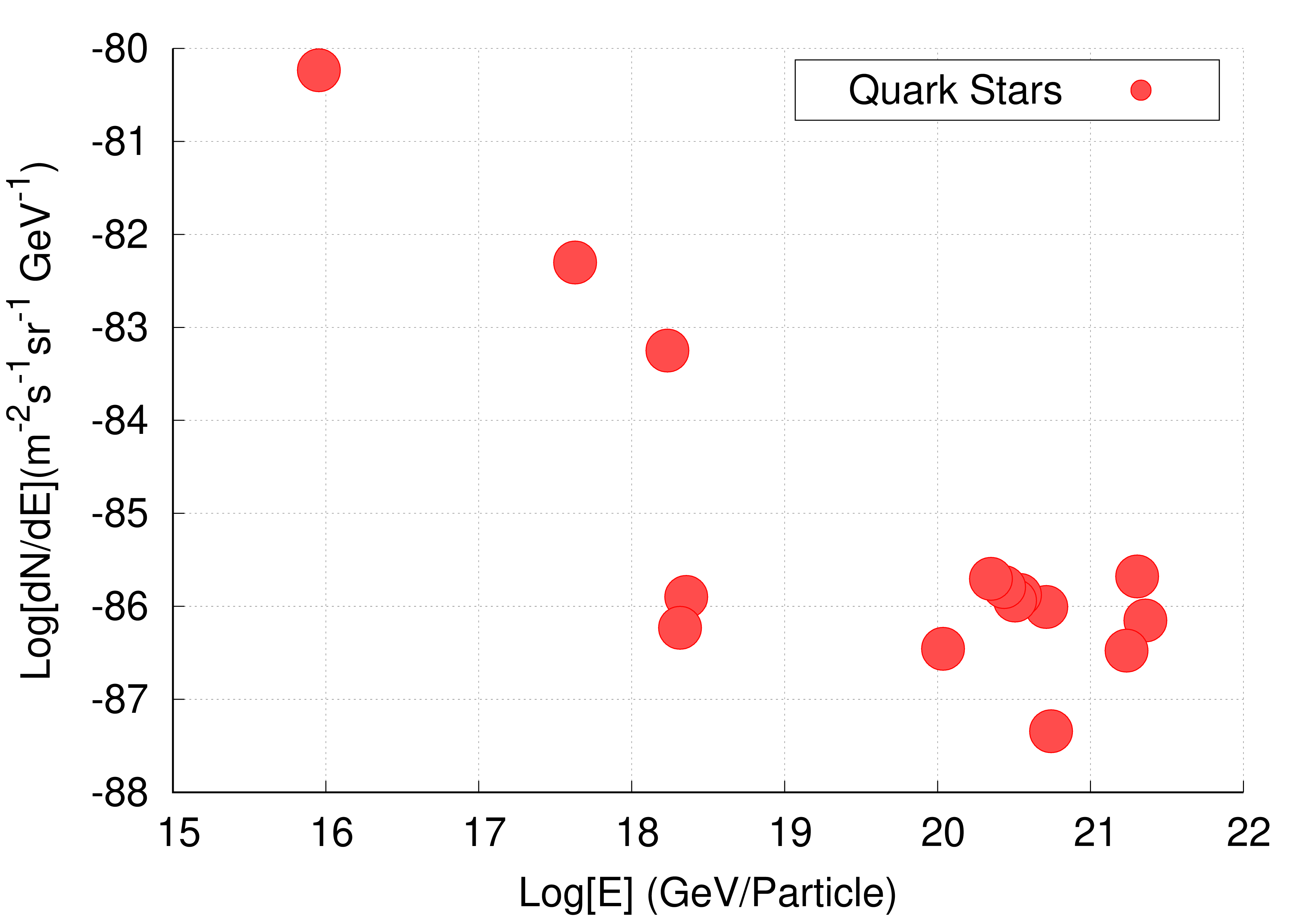}
    \caption{Log plot of the differential flux with respect to the energy per particle reaching earth. This flux obtained depends on both the distance distribution and the extreme spin vs. magnetic strength distribution. The flux of particles increases in this case where all the pulsars are taken to be quark stars.}
    \label{flux2}
\end{figure}

The differential flux is calculated for such distributions and is shown in figure \ref{flux1} and figure \ref{flux2}. In fig \ref{flux1}, the extra-galactic sources are taken to be quark stars, and the galactic sources are taken to be hadronic stars. Also, the magnetic field and rotational velocity of the stars are taken from the distribution shown in fig \ref{dis} (a). Fig \ref{flux2} shows the flux as a function of energy with the assumption that all the stars are quark stars and having high rotational and magnetic field values. The plots are shown in logarithmic scale, and converting to the normal scale the highest energy that the strangelets we detect are of the order of $10^9 - 10^{10}$ GeV.

The high energy strangelets flux expected on the earth lies on the interface of the galactic and extra-galactic high energy particles detected in the cosmic rays. Such high energy particles in cosmic rays is still a mystery as their charge to mass ratio is difficult to measure. However, the acceleration mechanism reported in this calculation shows that massive strange nuggets can be one of the candidates for such high energy particles.

The differential flux depends both on the distance distribution of the pulsars (across the galactic and extra-galactic regions) and the number of pulsars being considered. We see that as we change the distribution of pulsars, the differential flux gets shifted, and we get more flux of particles for the same energy. If a higher number of pulsars are taken into consideration, then the flux will be more.

\section{summary and discussion}
To summarize, in this work, we propose that the high rotational and magnetic field in pulsars can be the acceleration mechanism for the particles emitted from the surface of the star. The positive particles are mostly extracted near the equatorial region and negative particles from the polar region. We have also shown that the kinetic energy of the emitted particles is directly proportional to the surface magnetic field of the star, its rotational velocity, and also the mass of the emitting particle.

As the energy is directly proportional to the mass of the emitting positively particles, we focus our attention on the massive quark nuggets that can be emitted from the pulsars. Taking mostly the detected pulsars and assuming a given distribution \cite{andreasyan}, we have calculated the flux of the strangelets reaching the earth. We found that the flux could reach the "second knee" for nuclei, and up to the "ankle" for ordinary strangelet, in the cosmic ray energy spectrum. Therefore, this acceleration mechanism and the strangelets can be one of the candidates for high energy particles detected in cosmic ray showers.

We should mention that we have taken into account only the detected pulsars. However, the actual number of pulsars that can be present in the universe may be more by a few orders of magnitude. Therefore, our estimate of flux can also increase, however as the exact number of pulsars is not known, we refrain from such calculation. The acceleration mechanism of both positive and negative particles can also give rise to more interesting observational aspects, and our present efforts are towards exploring such avenues.

\begin{acknowledgments}
RB and RM wishes to acknowledge Project No. SB/ S2/ RJN-061/ 2015 Dated 25/05/2016 for financial support and thank Dr. Arunava Bhadra, University of North Bengal for useful suggestions.

\end{acknowledgments}

\appendix
\section{Radius correction due to rotation and magnetic field}
$$\Delta R_{r}(\theta)=(h_0^R(R_{s})-\Big(\frac{R^2_{s}\bar{\omega}^2e^{-\nu}}{3}\Big))$$
$$(\rho_0(R_{s})+p_0(R_{s}))\frac{\partial r}{\partial p_0(r)}\Big|_{R_{s}}$$
\begin{equation}
\Big[(h_2^R(R_{s})+\Big(\frac{R^2_{s}\bar{\omega}^2e^{-\nu}}{3}\Big))$$
$$(\rho_0(R_{s})+p_0(R_{s}))\frac{\partial r}{\partial p_0(r)}\Big|_{R_{s}}+R_{s}k_2^R(R_{s})\Big]P_2(\cos\theta)
\end{equation}

and

$$\Delta R_{m}(\theta)=(h_0^M(R_{s})-(\frac{2a_\phi j_\phi}{3R_{s}^2(\rho_0(R_{s})+p_0(R_{s}))}))$$
$$(\rho_0(R_{s})+p_0(R_{s}))\frac{\partial r}{\partial p_0(r)}\Big|_{R_{s}}$$
\begin{equation}
+\Big[(h_2^M(R_{s})+(\frac{2a_\phi j_\phi}{3R_{s}^2(\rho_0(R_{s})+p_0(R_{s}))}))$$
$$(\rho_0(R_{s})+p_0(R_{s}))\frac{\partial r}{\partial p_0(r)}\Big|_{R_{s}}+R_{s}k_2^M(R_{s})\Big]P_2(\cos\theta)
\end{equation}
\section{Expression for the Area}
$$
A=\zeta_2^R(0.30625 r k_2^M+0.30625 r k_2^R-0.75\zeta_0^M-0.75\zeta_0^R+0.30625\zeta_2^M$$
$$-0.75 r)+rk_2^M(0.30625 r k_2^R-0.75\zeta_0^M-0.75\zeta_0^R+0.30625\zeta_2^M$$
$$-0.75r)+0.153125 r^2(k_2^M)^2+0.153125 r^2(k_2^R)^2+rk_2^R(-0.75\zeta_0^M$$
$$-0.75\zeta_0^R+0.30625\zeta_2^M-0.75 r)+r^2+2\zeta_0^M\zeta_0^R-0.75 \zeta_0^M\zeta_2^M$$
$$+2r\zeta_0^M+(\zeta_0^M)^2-0.75\zeta_0^R\zeta_2^M+2r\zeta_0^R$$
\begin{equation}
	+(\zeta_0^R)^2-0.75r\zeta_2^M+0.153125(\zeta_2^M)^2+0.153125(\zeta_2^R)^2
\end{equation}
Where 
$$r=R_{s}$$
$$\zeta_0^M=(h_0^M(R_{s})-(\frac{2a_\phi j_\phi}{3R_{s}^2(\rho_0(R_{s})+p_0(R_{s}))}))$$
$$(\rho_0(R_{s})+p_0(R_{s}))\frac{\partial r}{\partial p_0(r)}\Big|_{R_{s}}$$
$$\zeta_2^M=h_2^M(R_{s})+(\frac{2a_\phi j_\phi}{3R_{s}^2(\rho_0(R_{s})+p_0(R_{s}))}))$$
$$(\rho_0(R_{s})+p_0(R_{s}))\frac{\partial r}{\partial p_0(r)}\Big|_{R_{surafce}}$$
$$\zeta_0^R=h_0^R(R_{s})-\Big(\frac{R^2_{s}\bar{\omega}^2e^{-\nu}}{3}\Big))$$
$$(\rho_0(R_{s})+p_0(R_{s}))\frac{\partial r}{\partial p_0(r)}\Big|_{R_{s}}$$
$$\zeta_2^R=h_0^R(R_{s})+\Big(\frac{R^2_{s}\bar{\omega}^2e^{-\nu}}{3}\Big))$$
$$(\rho_0(R_{s})+p_0(R_{s}))\frac{\partial r}{\partial p_0(r)}\Big|_{R_{s}}$$


\end{document}